\documentclass[letterpaper]{article}
\usepackage{xcolor}

\usepackage{flairs}

\usepackage{times}
\usepackage{helvet}
\usepackage{courier}
\usepackage{graphicx}
\usepackage{url}
\usepackage{amsmath,amssymb,amsfonts}
\usepackage{textcomp}
\usepackage{accessibility}
\begin{document}
\title{Prediction of Solar Flares Using Photospheric Magnetic Field Parameters with Deep Learning}
\author{
\hspace*{+3cm}Yash Chaudhary, Jason T. L. Wang, Chunhui Xu, Yan Xu\\
\hspace*{+3cm}New Jersey Institute of Technology \\
\hspace*{+3cm}Newark, NJ 07102, USA \\
\And 
\hspace*{+3.3cm}Sen Zhang \\
\hspace*{+3.3cm}State University of New York\\
\hspace*{+3.3cm}Oneonta, NY 13820, USA\\
}
\maketitle
\begin{abstract}
\begin{quote} 
Solar flares, particularly those of the M- and X-class,
have a significant impact on human life because of their potential to disrupt critical infrastructure and communication systems on Earth. Accurate prediction of solar flares is crucial for mitigating these risks, but the black-box nature of conventional deep learning models used in flare prediction limits their trustworthiness and interpretability.
In this paper, we propose a new approach to solar flare prediction using photospheric magnetic field parameters or features with deep learning.
To improve model interpretability, 
we integrate explainable artificial intelligence (XAI) techniques, 
including SHapley Additive exPlanations (SHAP) and partial dependence plots (PDPs),
into our prediction framework. 
XAI methods provide transparency by analyzing the importance and interactions of features used by 
our model.
Specifically, SHAP values offer a global 
and local understanding of the features, while PDPs 
provide insights into feature-level trends. 
These techniques demonstrate the potential of XAI in deploying AI-driven solutions in high-impact applications such as solar flare prediction, paving the way for more informed decision-making in solar physics and space weather studies.
\end{quote}
\end{abstract}

\section{Introduction}

Solar flares are intense bursts of radiation on the Sun's surface, typically caused by the release of magnetic energy in the Sun's atmosphere. They are classified into five categories based on their intensity: A, B, C, M, and X, with the X-class flares being the most powerful. Although smaller flares are generally harmless, larger M- and X-class flares can lead to coronal mass ejections that cause geomagnetic storms. These storms pose significant risks to critical infrastructure such as communication networks, satellite systems, and power grids, as well as radiation threats to astronauts and high-altitude airline passengers. 
Accurate solar flare prediction is crucial to mitigating these potential risks.
Predictive models 
typically analyze photospheric vector magnetograms, which reveal the magnetic field structure on the Sun's surface. 
Key features extracted from the magnetic field structure, such as magnetic flux and energy dissipation, 
can then be used to forecast the occurrence of solar flares.

In this work, we propose
a new deep learning model for the prediction of solar flares.
The effectiveness of our model is optimized through data preprocessing and feature selection techniques, including 
Analysis of Variance (ANOVA) 
\cite{guyon2003introduction} 
and 
Mutual Information (MI) 
\cite{peng2005feature},
which together produce an optimal subset of features for the learning process.
Furthermore, despite the strong predictive performance of the 
proposed model, 
its black-box nature poses challenges to interpretability. 
To address this, we incorporate explainable artificial intelligence (XAI) techniques, 
including SHapley Additive exPlanations (SHAP) 
\cite{lundberg2017unified}
and
partial dependence plots (PDPs)
\cite{molnar2020interpretable},
into
our prediction framework. 
These XAI methods provide transparency into
our model's decision-making process, helping to identify the importance and interactions of features, while also offering explanations for individual predictions. This ensures that the model output is not only accurate, but also interpretable.
In what follows, we first
present the dataset and data preprocessing methods used in our study.
Then, we
describe  
our approach in detail.
Next, 
we delve into the integrated XAI techniques.
Finally, we conclude the paper and point out some directions for future research.

\subsection{Related Work}
Recent advancements in time-series forecasting (TSF) have heavily leveraged transformer architectures to capture long-range dependencies. 
Foundational works introduced self-attention mechanisms to TSF
\cite{VaswaniSPUJGKP17},
while subsequent models such as Informer 
\cite{DBLP:conf/aaai/ZhouZPZLXZ21}
addressed computational bottlenecks in long sequences. More recently, models like PatchTST 
\cite{DBLP:conf/iclr/NieNSK23}
demonstrated that segmenting time series into discrete patches significantly improves both forecasting accuracy and computational efficiency, and introduced the idea of channel-independent modeling. Our work adapts these state-of-the-art patching concepts to the specific domain of solar flare prediction. In the solar physics domain, time-series feature selection for solar 
eruption prediction has also been explored 
\cite{AJLL2022,2023NatSR..1313665A,2024ApJ...961...81F,2025ApJ...981...37Z},
providing further motivation for our feature selection pipeline.

\section{Data Preprocessing and Preparation}

Our research utilizes two key data sources 
for solar flare prediction: 
the Space-weather HMI Active Region Patches (SHARPs) data products 
and integrated Lorentz force estimates \cite{liu2019predicting}. 
The two primary data series we used are the {\tt hmi.sharp} series and the {\tt cgem.Lorentz} series, which were queried from the Joint Science Operations Center (JSOC) website 
(\url{http://jsoc.stanford.edu/})
using SunPy \cite{sunpy2015}.
The {\tt hmi.sharp} series provides automatically identified and tracked active regions in solar map patches, including crucial physical parameters for flare prediction
\cite{2014SoPh..289.3549B}. 
The {\tt cgem.Lorentz} series offers estimates of the integrated Lorentz forces, which help us to explore the dynamic processes within each active region
\cite{2012SoPh..277...59F}.
By studying these forces, we gain insight into the accumulation and release of magnetic energy, 
a key factor in flare activity.

To train our deep learning model, we created a data set consisting of M- and X-class flares and their corresponding active regions, collected from May 2010 to May 2018. This data set was derived from the Geostationary Operational Environmental Satellite (GOES) X-ray flare catalog provided by the National Centers for Environmental Information (NCEI). The time series data set of the physical parameters, collected at a 1 hour frequency, allowed us to train the proposed model using the parameters of the active regions and the corresponding Lorentz forces during the study period.
Specifically,  
we focused on the 25 photospheric magnetic field parameters suggested in the literature \cite{bobra2015solar} as the initial set of predictive features. 

Given that not all 25 features contribute equally to the prediction task, feature selection plays a vital role in improving the performance of the proposed model by eliminating redundant or irrelevant features.
We employed two feature selection methods:
ANOVA (Analysis of Variance) \cite{guyon2003introduction} and
MI (Mutual Information) \cite{peng2005feature}. 
These methods were used to reduce the number of features and focus on the most relevant ones for the prediction task. 
The scores of both methods were normalized using Min-Max scaling in a range of 0 to 1 before applying the thresholds of the methods.
The ANOVA F-scores and MI scores were calculated using 
only training data. This is crucial to prevent data leakage
from the test set in calculating the feature scores. 
The optimal thresholds for these scores were then determined using Optuna \cite{akiba2019optuna} to maximize the true skill statistic 
on the test set.

ANOVA F-scores were calculated to assess the importance of each of the 25 magnetic features by examining the variance between classes.
Features with a normalized ANOVA F-score higher than or equal to a threshold of 0.1 were
selected where the optimal threshold was determined using Optuna as mentioned above; 
this resulted in 12 features meeting this criterion.
MI scores were calculated to measure the amount of information shared between each feature and the target variable. For this study, features with a normalized MI score higher than or equal to a 
threshold of 0.2 were selected
where the optimal threshold was also determined using Optuna;
this resulted in 15 features meeting this criterion.
After performing these individual selections based on normalized ANOVA and MI scores,
we took the union of the set of the 12 features selected by the normalized ANOVA F-score 
and the set of the 15 features selected by the normalized MI score to obtain
the final set of features used in this study. 
This combined set, which represents the features deemed important by at least one of the methods, 
contains 16 features, as shown in Table \ref{tab:features}.

\begin{table}[t!]
\alt{All Selected 16 Features and Their Descriptions}
\caption{All Selected 16 Features and Their Descriptions}
\label{tab:features} 
\begin{center}
\begin{small}
\begin{tabular}{ll}
\hline
\textbf{Keyword} & \textbf{Description} \\ \hline
ABSNJZH & Absolute value of the net current helicity \\ \hline
AREA-ACR & Area of strong field pixels in active region \\ \hline
MEANALP & Mean characteristic twist parameter, $\alpha$ \\ \hline
MEANJZH & Mean current helicity \\ \hline
MEANPOT & Mean photospheric magnetic free energy \\ \hline
MEANSHR & Mean shear angle \\ \hline
R-VALUE & Sum of flux near polarity inversion line \\ \hline
SAVNCPP & Sum of the modulus of the net current  \\ \hline
TOTBSQ & Total magnitude of Lorentz force \\ \hline
TOTFX & Sum of x-component of Lorentz force \\ \hline
TOTFY & Sum of y-component of Lorentz force \\ \hline
TOTFZ & Sum of z-component of Lorentz force \\ \hline
TOTPOT & Total  magnetic energy density \\ \hline
TOTUSJH & Total unsigned current helicity \\ \hline
TOTUSJZ & Total unsigned vertical current \\ \hline
USFLUX & Total unsigned flux \\ \hline
\end{tabular}
\end{small}
\end{center}
\end{table}

\begin{table}[t!]
\alt{Data Distribution in Our Study}
\caption{Data Distribution in Our Study}
\label{tab:data}
\begin{center}
\begin{small}
\begin{tabular}{lcc}
\hline
\textbf{Dataset} & \textbf{Positive Sample No.} & \textbf{Negative Sample No.} \\
\hline
Training  & 2,157 & 65,504 \\
Test & 553 & 16,363 \\
\hline
\end{tabular}
\end{small}
\end{center}
\end{table}

\section{Our Approach}
\label{sec:method}

Each record at time step $t$ contains 16 features, shown in Table \ref{tab:features}.
We label the record as positive (class 1) indicating that there will be 
a $\geq$M-class flare
(i.e., an M- or X-class flare)
within the next 24 hours of $t$, or as negative (class 0) indicating
that there will be no $\geq$M-class flare
within the next 24 hours of $t$.
To let our model learn the evolution of magnetic fields,
we construct samples, where each sample is a time series or sequence
of $T = 24$ records.
The label of the sample/sequence is
defined as the label of the last record in the sequence.
Samples containing missing or corrupted records are excluded from the study.
To explicitly address the inherent class imbalance in the data sets, we utilize a stratified training/test split strategy, which
preserves the original proportion of each class (positive vs. negative) within both 
training and test sets. 
The distribution of target classes (positive vs. negative) in the training and test sets is shown in Table \ref{tab:data}.
The training set contains
67,661 samples, while the test set
contains 16,916 samples. 

The primary task is to predict binary outcomes (positive vs. negative)
based on input samples. 
The goal is to classify each test sample  
into one of two classes, with the prediction being positive (class 1) or negative (class 0).
When a test sample ending at the time step $t$ is predicted to be positive, 
it means there will be a $\geq$M-class flare within the next 24 hours of $t$. 
If the test sample is predicted to be negative,
it means there will be no $\geq$M-class flare within the next 24 hours of $t$. Thus, we essentially use the magnetic field parameters or features 
in the 24 hours 
before $t$ to predict the flare occurrence in the 24 hours after $t$.

\begin{figure*}[ht!]
    \centering
    \includegraphics[width=2.00\columnwidth]{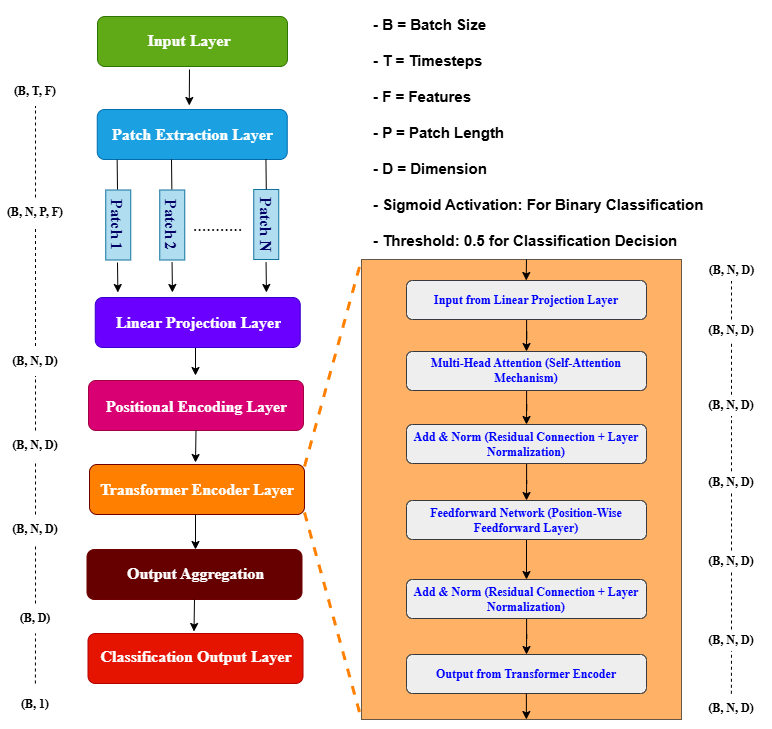} 
    \alt{Architecture of the proposed deep learning model for solar flare prediction.}
    \caption{Architecture of the proposed deep learning model for solar flare prediction.}
    \label{fig:MTST_architecture} 
\end{figure*}

\subsection{The Proposed Model}
Figure \ref{fig:MTST_architecture} presents the architecture of the
proposed deep learning model. 
Our model processes samples by first dividing the input
sample into smaller, non-overlapping patches.
The model employs
self-attention mechanisms to capture both local and global dependencies across patches of the input sample. The primary components of the architecture are:

\begin{itemize}  
    \item Input Layer: Takes an input sample with \( T \) time steps and \( F \) features, represented as a tensor of shape \( (B, T, F) \), where \( B=64 \) is the batch size, \( T=24 \) is the number of time steps, and \( F=16 \) is the number of features.
    
    \item Patch Extraction Layer: Divides the input sequence into non-overlapping patches of fixed length \( P=2 \) and dimensionality \( F=16 \), resulting in patches of shape \( (B, N, P, F) \), 
    where \( N = \frac{T}{P} = 12 \) is the number of patches.
    
    \item Linear Projection Layer: Projects each patch into a higher-dimensional space with dimensionality \( D = 128 \), transforming the patches 
    into shape \( (B, N, D) \). This projection allows the model to better capture the temporal dependencies within each patch.
    
    \item Positional Encoding Layer: Adds positional information to the patches to ensure that the model retains the temporal order of the time series data. 
    The shape remains \( (B, N, D) \).
    
    \item Transformer Encoder Layer: 
    Uses multiple self-attention layers (with multi-head attention) and feedforward networks to capture local and global dependencies within patches, keeping the output shape as \( (B, N, D) \).
    
    \item Output Aggregation (Flatten + Linear Head): The output of the Transformer Encoder 
    Layer is aggregated by flattening the patch-wise representations and passing them through a linear layer. This results in a tensor of shape \( (B, D) \), representing aggregated features.
    
    \item Classification Output Layer: A fully connected layer followed by a sigmoid activation function is applied to produce the final classification probability. The output is a tensor of shape \( (B, 1) \), where each 
    test sample 
    corresponds to a binary class prediction.
\end{itemize}

The proposed model predicts binary class labels 
based on the extracted patch-level features. 
Mathematically, this is expressed as follows:
\[
P(\hat{y} = 1) = \sigma(W \cdot \text{Patch Features} + b),\]
where
    \( \sigma \) is the sigmoid activation function,
    \( W \) represents learned weights, and
    \( b \) is the bias term.

The model is trained using the
weighted cross-entropy loss
\cite{liu2019predicting}, 
with more weight given to the minority (i.e., positive)
class.
A decision threshold of 0.5 is used to classify the output.  
    If \( P(\hat{y} = 1) \geq 0.5 \), i.e.,
    if the predicted probability for a test sample to be positive
    is greater than or equal to 0.5,
    then the test sample is classified as positive (class 1).  
    Otherwise, the test sample is classified as negative (class 0).  

\subsection{Model Evaluation}
The performance of our model was evaluated using multiple metrics. Given the highly imbalanced nature of
our
solar flare dataset (see Table \ref{tab:data})—where a naive majority-class classifier could superficially achieve over 96\% accuracy by always predicting the negative class—we emphasize the 
%true skill statistic (TSS) 
area under curve (AUC)
and recall as our primary evaluation metrics. 
The raw test accuracy is reported for completeness only. 
The results are as follows:

\begin{itemize}
    \item Test Accuracy: The model achieved a test accuracy of 95.18\%.
    \item Recall: The recall was 0.89, demonstrating the effectiveness of the model in identifying positive samples.
    \item Area Under Curve (AUC): The AUC was 0.96, reflecting excellent classification performance.
\end{itemize}

Our model outperforms a state-of-the-art method 
\cite{2025ApJS..279...27H},
and surpasses the best-performing Transformer model 
\cite{2025ApJS..276....7L}.
It is important to note the differences in the experimental configurations when comparing these baselines: 
Li et al. (2025) 
utilized a 24-hour observation window with a 36-minute sampling frequency (sequence length of 40), while our model operates on a 1-hour frequency (sequence length of 24), demonstrating robust predictive capability even with 
shorter sequences.

\section{XAI Techniques for Interpretability}
\label{sec:xai}

To ensure interpretability and transparency of our model's predictions, we employed XAI techniques (SHAP and PDPs), which provided insights into feature importance, individual predictions, and feature dependencies, supported by comprehensive visualizations. 

\subsection{SHAP Results}

SHAP values quantify the contribution of each feature to the model's predictions, offering a robust framework for interpretability. By explaining both global feature importance and local prediction behavior, SHAP enhances our understanding of
the proposed model's decision-making process. 
Below are detailed visualizations generated using SHAP 
for the solar flare prediction task.

\subsubsection{Feature Importance Plot}
\begin{figure}
    \centering
    \includegraphics[width=0.48\textwidth]{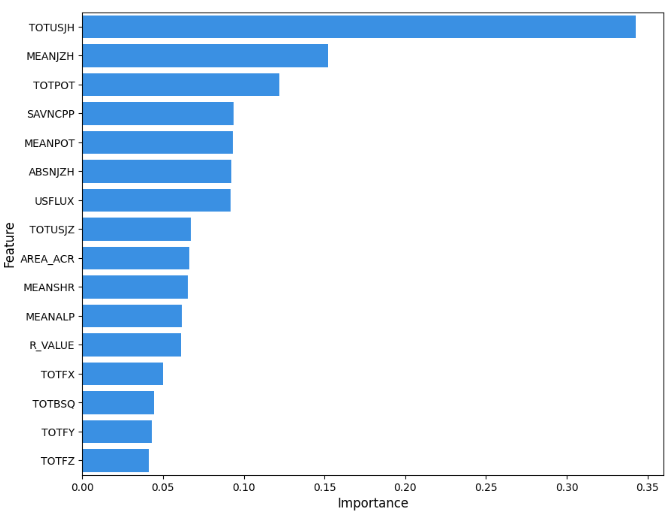} 
    \alt{SHAP feature importance plot: Bar graph ranking the features or predictors based on their overall impact on the model's predictions. Features higher on the list have a greater influence.}
    \caption{SHAP feature importance plot: Bar graph ranking the features or predictors based on their overall impact on the model's predictions. Features higher on the list have a greater influence.}
    \label{fig:shap_importance}
\end{figure}

The feature importance plot ranks the features or
predictors based on their importance,
calculated by mean absolute SHAP values, which reflect their overall contribution to the model's predictions (Figure~\ref{fig:shap_importance}). This visualization aids in identifying key features driving the model's behavior, such as TOTUSJH, MEANJZH, and TOTPOT. 

\subsubsection{Beeswarm Plot}
\begin{figure}[t!]
    \centering
    \includegraphics[width=0.48\textwidth]{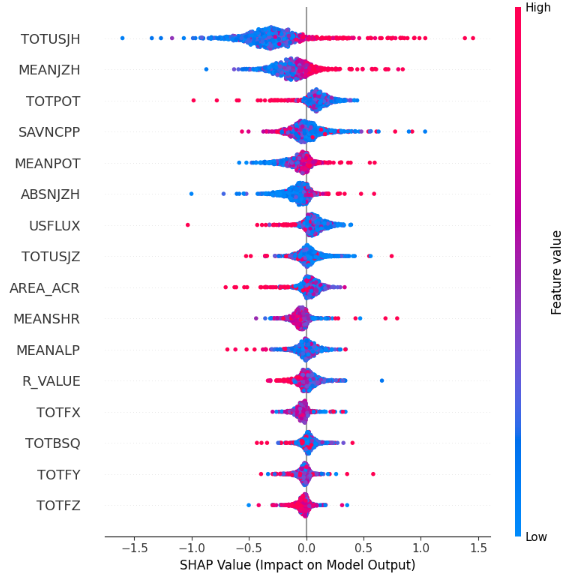}  
    \alt{SHAP beeswarm plot: Features ranked by importance, with SHAP value distribution visualized to show the impact on individual predictions. Each dot represents an individual prediction corresponding to a specific test sample, color-coded by feature value (blue to red).}
    \caption{SHAP beeswarm plot: Features ranked by importance, with SHAP value distribution visualized to show the impact on individual predictions. Each dot represents an individual prediction corresponding to a specific test sample, color-coded by feature value (blue to red).}
    \label{fig:shap_beeswarm}
\end{figure}

The beeswarm plot provides a detailed view of SHAP values for each feature across the test set (Figure~\ref{fig:shap_beeswarm}). It visualizes how features contribute positively or negatively to predictions, with color coding indicating the feature value. 
For instance, for the
TOTUSJH feature, many high feature values (in red)
contribute positively, with positive SHAP values, to the model's predictions.
Many low feature values (in blue) contribute negatively, with negative SHAP values, 
to the model's predictions. 

\begin{figure}[t!]
    \centering
    \includegraphics[width=0.445\textwidth]{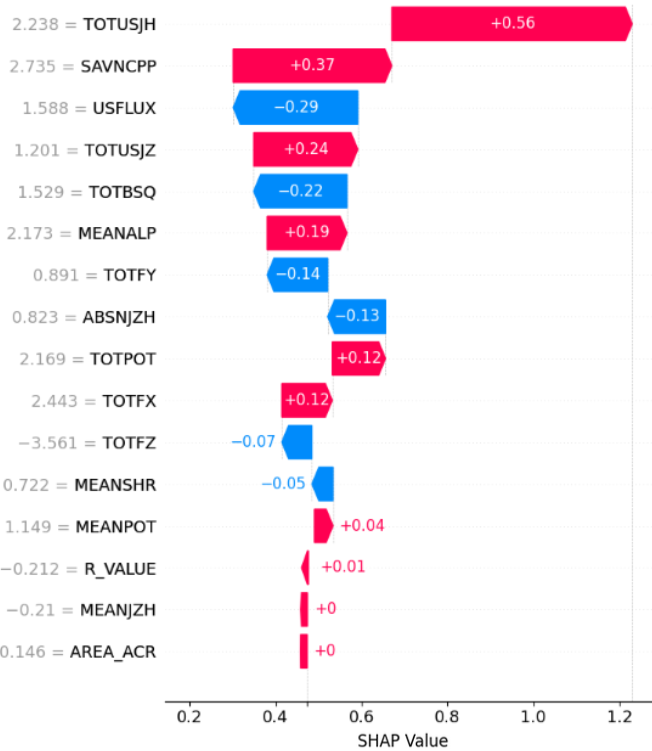} 
    \alt{SHAP waterfall plot: Cumulative feature contributions for a specific test sample predicted to be positive (class 1). 
    The plot illustrates how contributions from individual features accumulate to determine the final prediction. TOTUSJH exhibits the highest positive impact (with a SHAPE value of $+0.56$), while USFLUX has the highest negative impact (with a SHAP value of $-0.29$).}
    \caption{SHAP waterfall plot: Cumulative feature contributions for a specific test sample predicted to be positive (class 1). 
    The plot illustrates how contributions from individual features accumulate to determine the final prediction. TOTUSJH exhibits the highest positive impact (with a SHAPE value of $+0.56$), while USFLUX has the highest negative impact (with a SHAP value of $-0.29$).} 
    \label{fig:shap_waterfall}
\end{figure}

\begin{figure}[t!]
    \centering
    \includegraphics[width=0.445\textwidth]{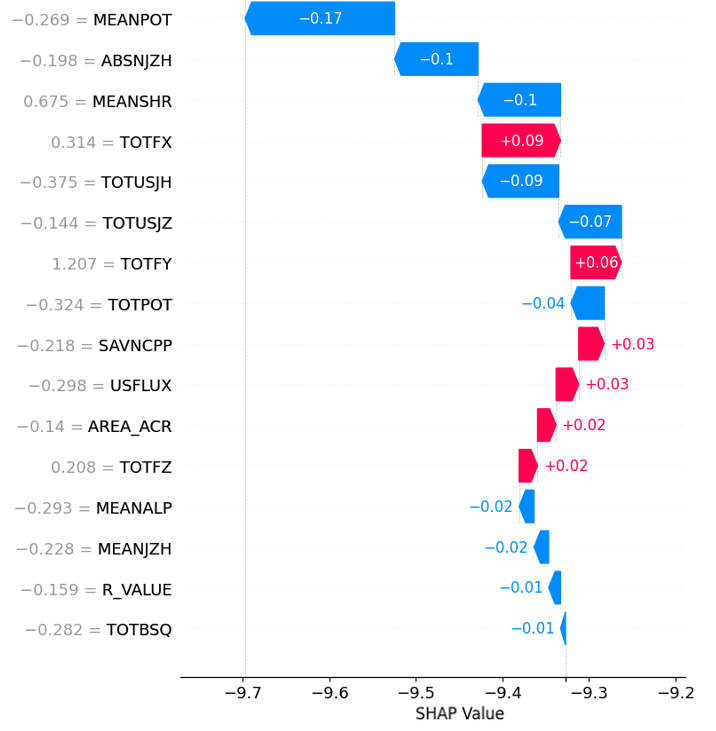} 
    \alt{SHAP waterfall plot: Cumulative feature contributions for another test sample, predicted as negative (class 0). 
    For this test sample, TOTUSJH (with a SHAP value of $-0.56$) and SAVNCPP (with a SHAP value of $-0.37$) provide strong contributions towards the class 0 prediction, while USFLUX has the largest contribution (with a SHAP value of 
    $+0.29$) in pushing the prediction away from the class 0.}
    \caption{SHAP waterfall plot: Cumulative feature contributions for another test sample, predicted as negative (class 0). 
For this test sample, MEANPOT (with a SHAP value of $-0.17$) and ABSNJZH (with a SHAP value of $-0.10$) provide strong contributions towards the class 0 prediction, while TOTFX (with a SHAP value of $+0.09$) has the largest contribution in pushing the prediction away from class 0.}
    \label{fig:shap_waterfall_negative}
\end{figure}

\subsubsection{Waterfall Plot}
The waterfall plot explains the prediction of the model for a specific test sample by visualizing the cumulative SHAP values.
Figure~\ref{fig:shap_waterfall} presents the waterfall plot for
a test sample, which is predicted to be positive (class 1).
In Figure~\ref{fig:shap_waterfall},
the positive and negative contributions of features such as TOTUSJH and USFLUX are clearly delineated, showcasing their role in the prediction for class 1.

Figure~\ref{fig:shap_waterfall_negative} shows the waterfall plot for another test sample, which our model predicts as negative (class 0).
This plot illustrates how features such as MEANPOT, ABSNJZH, and MEANSHR strongly contribute to this negative prediction with negative SHAP values, while features such as TOTFX and TOTFY push the prediction away from class 0 with positive SHAP values.
Note that the feature rankings in the waterfall plots,
which provide local explanations for specific test samples,
are different from the feature rankings
in the feature importance plot and beeswarm plot, 
which provide global explanations with respect to the entire test set.

\subsection{PDP Results}

Partial dependence plots (PDPs) provide insights into the marginal effects of one or more features on the model's predictions.
By analyzing PDPs, we can observe the trend and interaction at the feature level.
We consider 2-dimensional (2D) PDPs and
3-dimensional (3D) PDPs here.

\subsubsection{2D PDP}
\begin{figure}[t!]
    \centering
    \includegraphics[width=0.95\columnwidth]{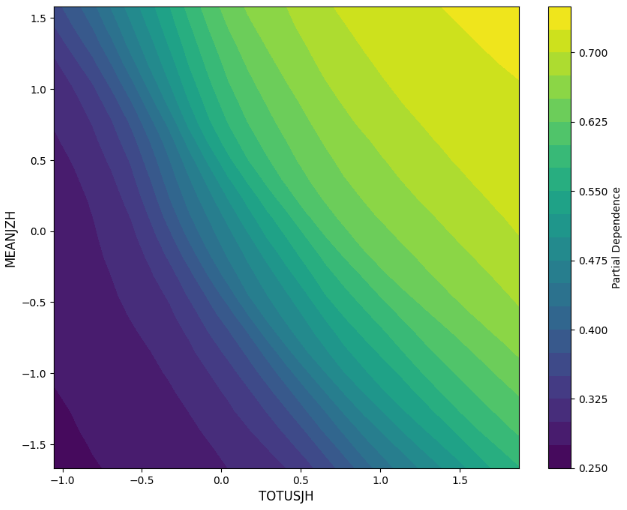}
    \alt{2D PDP showing the interaction between TOTUSJH and MEANJZH. Color, as indicated by the color bar, represents the average predicted probability (partial dependence).}
    \caption{2D PDP showing the interaction between TOTUSJH and MEANJZH. Color, as indicated by the color bar, represents the average predicted probability (partial dependence).}
    \label{fig:pdp_2d_interaction}
\end{figure}

A 2D PDP is used to visualize the interaction effect between two features on the model's predictions. Figure~\ref{fig:pdp_2d_interaction} presents the
2D PDP for the TOTUSJH and MEANJZH features.
The plot is displayed as a heatmap, where the color intensity, indicated by the color bar, represents the
average predicted probability (partial dependence). 
In this plot, warmer colors (i.e., yellow) signify higher average predicted probabilities, while cooler colors (i.e., dark purple/blue) indicate lower probabilities. 
This visualization shows that the highest predicted probabilities (up to $\approx 0.7$) occur when both TOTUSJH and MEANJZH have high values (e.g., $>$ 1.0). When TOTUSJH is low, the average predicted probability remains low even if MEANJZH is high, suggesting that TOTUSJH has a dominant effect.

\subsubsection{3D PDP}
\begin{figure}
%[ht!]
    \centering
    \includegraphics[width=0.95\columnwidth]{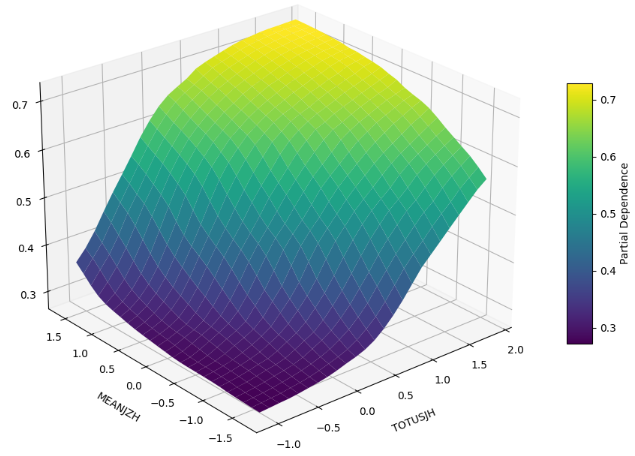}
    \alt{3D PDP providing enhanced visualization of the interaction effects between TOTUSJH and MEANJZH on the model's predictions. The color bar indicates the average predicted probability corresponding to the surface color and height.}
    \caption{3D PDP providing enhanced visualization of the interaction effects between TOTUSJH and MEANJZH on the model's predictions. The color bar indicates the average predicted probability corresponding to the surface color and height.}
    \label{fig:pdp_3d_interaction} 
\end{figure}

To further understand the interaction effects
between TOTUSJH and MEANJZH, a 3D PDP 
is generated (Figure~\ref{fig:pdp_3d_interaction}). This plot visualizes their combined effects on the model's predictions, with the $x$-axis representing TOTUSJH, the $y$-axis representing MEANJZH, and the $z$-axis (height of the surface) indicating the average predicted probability (partial dependence). 
The surface is color-coded according to the $z$-axis values (average predicted probability), as detailed by the color bar.
For instance, the yellow/brighter regions on the surface correspond to higher predicted probabilities, while dark purple/blue regions indicate lower probabilities. The 3D visualization reinforces the interaction seen in the 2D heatmap, showing a complex surface where, for example, the predicted probability increases with increasing
TOTUSJH values. 
The surface shows a clear gradient, illustrating that 
both TOTUSJH and MEANJZH features contribute to the prediction.

The integration of XAI techniques in this study has 
improved the interpretability of the proposed model's predictions for solar flare forecasting. By employing SHAP and PDPs, we have gained comprehensive insight into feature importance, interaction effects, and individual prediction behaviors. These visualizations and analyses not only reinforce model transparency but also ensure that the decision-making process is understandable and grounded in the underlying data, marking a crucial step towards reliable, trustworthy and explainable AI solutions.

\section{Conclusion}
\label{sec:conclusion}

In this paper, we present a new deep learning model for solar flare prediction, addressing a critical challenge in space weather forecasting. By carefully selecting relevant features and employing explainability techniques, such as SHAP and PDPs, we have demonstrated the interpretability of our model's predictions. The results show that the model coupled with the chosen metrics---accuracy, recall and AUC---provides a robust framework for solar flare forecasting. Incorporation of XAI methods, namely SHAP and PDPs, not only enhances transparency, but also promotes trust 
in AI-driven decision-making processes.
Future work includes
\begin{itemize}
\item
exploration of advanced XAI techniques  
such as counterfactual explanations 
and integrated gradients 
to gain deeper insights into our model's predictions; 
\item 
development of a real-time solar flare prediction system leveraging the proposed model for operational use;
\item  
methodological expansions including investigation of variable prediction spans and sequence lengths to probe the model’s temporal sensitivity, ablation studies on reduced feature subsets (e.g., removing low-importance features such as AREA-ACR and R-VALUE) to assess their impact on performance and inference efficiency, and deeper statistical characterization of our dataset—such as autocorrelation analysis—to benchmark space weather time-series data against standard 
time-series forecasting
datasets.
\end{itemize}

\section{Acknowledgments}
We thank anonymous reviewers for their helpful comments and constructive suggestions.
This work was supported in part 
by NSF grants AGS-2228996 and RISE-2425602.
This support is greatly appreciated.

\bibliography{main}

\bibliographystyle{flairs}

\end{document}